\pdfoutput=1
\documentclass[journal=jpccck,etalmode=truncate,maxauthors=10,layout=twocolumn,manuscript=article]{achemso}
\setkeys{acs}{usetitle = true}
\usepackage{amssymb,amsmath}
\usepackage{graphicx}

\title{Controlled isotope arrangement in $^{13}$C enriched carbon nanotubes}

\author{J.~Koltai}
\email{koltai@elte.hu}
\author{G.~Mezei}
\affiliation[E{\"o}tv{\"o}s University]
{Department of Biological Physics, E{\"o}tv{\"o}s
University, P{\'a}zm{\'a}ny P{\'e}ter s{\'e}t{\'a}ny 1/A, H-1117
Budapest, Hungary}
\author{V.~Z{\'o}lyomi}
\affiliation[University of Manchester]
{National Graphene Institute, University of Manchester, Manchester M13 9PL, United Kingdom}
\author{J.~K{\"u}rti}
\affiliation
{Department of Biological Physics, E{\"o}tv{\"o}s University, P{\'a}zm{\'a}ny P{\'e}ter s{\'e}t{\'a}ny 1/A, H-1117 Budapest, Hungary}
\author{H.~Kuzmany}
\author{T.~Pichler}
\author{F.~Simon}
\affiliation[Universit{\"a}t Wien]
{Universit{\"a}t Wien, Fakult{\"a}t f{\"u}r Physik, Strudlhofgasse 4, 1090 Wien, Austria}
\alsoaffiliation[Budapest University of Technology and Economics]
{Department of Physics, Budapest University of Technology and Economics
and MTA-BME Lend{\"u}let Spintronics Research Group (PROSPIN), P.O. Box 91, H-1521 Budapest, Hungary}

\begin{document}

\begin{abstract}
We report the synthesis of a novel isotope engineered $^{13}{\rm C}$--$^{12}{\rm C}$ heteronuclear nanostructure: single-wall carbon nanotubes made of $^{13}{\rm C}$ enriched clusters which are embedded in natural carbon regions. The material is synthesized with a high temperature annealing from $^{13}{\rm C}$ enriched benzene and natural ${\rm C}_{60}$, which are co-encapsulated inside host SWCNTs in an alternating fashion. The Raman 2D line indicates that the 
$^{13}{\rm C}$ isotopes are not distributed uniformly in the inner tubes. A semi-empirical method based modeling of the Raman modes under $^{13}{\rm C}$ isotope enrichment shows that experimental data is compatible with the presence of $^{13}{\rm C}$ rich clusters which are embedded in a natural carbon containing matrix. This material may find applications in quantum information processing and storage using nuclear spins as qubits.
\end{abstract}

\section{Introduction}
Isotope substitution is a powerful tool to study phenomena which depend on the atomic mass. Examples where isotope substitution (also known as \textit{isotope labeling} or \textit{isotope engineering}) play a crucial role include identification of vibrational modes in ${\rm C}_{60}$ fullerene\cite{MihalyPRB1995,KendzioraPRB1995,HoroyskiPRB1996} and the evidence for phonon-mediated superconductivity \cite{BCS}. On the practical side, isotope engineering allows e.g. to perform nuclear magnetic resonance (NMR) on otherwise $I=0$ nuclei \cite{AbragamBook}, to study heat conduction in mononuclear materials \cite{CapinskiAPL1997}, and proposals exist for the use of such systems for nuclear quantum computing (QC) \cite{Shlimak,DivincenzoQC}. The latter proposal relies on the use of the nuclear spins as quantum-bits (or qubits) as nuclei couple weakly to the environment, and thus retain their quantum nature for longer times, but they can be effectively controlled using NMR schemes. 

Compelling realization of quantum computing would be a solid state device using a linear array of $^{31}{\rm P}$ doped silicon based heterostructures \cite{KaneQC} or a linear array of electron spin-qubits made of ${\rm N@C}_{60}$ \cite{HarneitPRA}. Common in both suggestions is the controlled arrangement of the working spin-qubit, which proves to be difficult in practice.

Isotope engineering is clearly desired for carbon allotropes, including fullerenes, single-wall carbon nanotubes (SWCNTs), and graphene, in order to exploit their QC and spintronics potential and to yield further insight into isotope dependent phenomena in these materials. An interesting example of smart isotope engineering is the growth of small diameter nanotubes inside host SWCNTs from isotope enriched ${\rm C}_{60}$'s with a high temperature annealing ($1200{\rm -}1300\,^{\circ}{\rm C}$) \cite{SimonPRL2005}. The resulting material is double-wall carbon nanotubes with natural carbon outer and highly $^{13}{\rm C}$ enriched inner walls. This material allowed the identification of the inner tube vibrational modes \cite{SimonPRL2005} and an inner tube specific NMR experiment \cite{SingerPRL2005}, which led to the NMR confirmation of the Tomonaga-Luttinger liquid state in carbon nanotubes \cite{SingerPRL2005,IharaEPL2010,DoraPRL2007}.

However, even for this material, isotope arrangement on the inner tube is random which is due to the random distribution of $^{13}{\rm C}$ nuclei over the starting ${\rm C}_{60}$ molecule \cite{SimonCPL2005_C59N}. Clearly, some level of control over the linear arrangement of $^{13}{\rm C}$ nuclei is desired for the reasons above. As a first step in this direction, the co-encapsulation of two types of ${\rm C}_{60}$'s with different enrichment levels was performed \cite{ZolyomiPRB2007}. However, the encapsulation proceeds with little or no particular preference for the two kinds of ${\rm C}_{60}$ molecules (diffusion constant is known to go as $\propto m^{-0.5}$, which gives a $4\%$ difference) and the resulting inner tubes showed further a random isotope distribution although with a somewhat different level of randomness \cite{ZolyomiPRB2007}. A further attempt can be made using different molecules which may enter in an ordered fashion. Isotope labeling of the two molecules could give rise to the desired 
linear arrangement of isotopes.

This co-encapsulation was explored using small organic molecules (benzene and toluene) together with ${\rm C}_{60}$ inside the host nanotubes and inner tubes were grown from them \cite{SimonCPL2006}. A more recent study\cite{ZolyomiJPCC2014} showed that the inner tube formation is preceded by the synthesis of larger, unconventional fullerenes (${\rm C}_{66}$ and ${\rm C}_{68}$) for a more moderate, $800\,^{\circ}{\rm C}$ annealing. For these intermediate products, the isotope enriched 6 membered rings were found to stay together, a property which might be inherited to the final inner tube material.

Motivated by the above findings, in this work we study the final inner tube product, which is made of benzene and ${\rm C}_{60}$, which are co-encapsulated inside the host nanotubes. We study the 2D Raman line, which is one of the most energetic phonon modes, and observe an unexpected change when the benzene is fully $^{13}{\rm C}$ enriched. Random distribution of the isotopes would cause a uniform downshift of the Raman mode, with its center corresponding to the $\sim 9\%$ isotope enrichment. In contrast, we observe a 2D Raman component which is downshifted as if it had an $85\%$ isotope enrichment. This is clearly incompatible with the random distribution of the different carbon isotopes and it suggests a clustering of $^{13}{\rm C}$. This is supported by a semi-empirical method based modeling of the Raman modes. Our calculations show that the Raman spectra can be explained by assuming the clusterization of the $^{13}{\rm C}$ nuclei: the $^{13}{\rm C}$ atoms form benzene-like rings on the inner nanotube 
wall which are separated from other such rings by natural carbon regions.

\section{Methods}

The SWCNT host sample was prepared by the arc-discharge method and was from the same batch as in previous studies \cite{SimonCPL2006} with a mean diameter of $1.4\,\rm nm$. This diameter is ideal for the growth of inner tubes as it can energetically well encompass the filled-in fullerenes. Prior to the filling process, the nanotubes were opened by heating in air at $450\,^{\circ}{\rm C}$ for 0.5 h. Fullerenes were obtained from a commercial source (Hoechst, Super Gold Grade ${\rm C}_{60}$, purity $99.9\%$) and we used natural (Sigma) and fully $^{13}{\rm C}$ enriched ($^{13}{\rm C}_{6}{\rm H}_{6}$, Eurisotop, France) benzene. We co-encapsulated the fullerenes and benzene by sonicating the SWCNTs for 2 hours in a benzene:${\rm C}_{60}$ solution of $1\,\rm mg/ml$. This is known to result in a clathrate structure where $\text{C}_6\text{H}_6$ and ${\rm C}_{60}$ molecules occupy 50\%-50\% of the available inner volume in an alternating fashion \cite{SimonCPL2006,ZolyomiJPCC2014}. The resulting filled nanotubes 
were removed from the solution by filtering and subsequently rinsed with an abundant benzene 
solvent to remove any non-encapsulated excess fullerenes from the outside of the host nanotubes. This was followed by the final filtering and drying of the resulting \textit{bucky-paper} samples under a fume hood. The samples were annealed in high dynamic vacuum at $1250\,^{\circ}{\rm C}$ for 1 hour. This method is known to yield high quality double wall carbon nanotubes \cite{SimonPRB2005,SimonCPL2006}. In the following, we denote double-wall carbon nanotubes grown from ${\rm C}_{60}$ and benzene both containing natural carbon as $^{12}{\rm DWCNT}$. $^{12/13}{\rm DWCNT}$ denotes double wall carbon nanotubes where the inner wall is made of co-encapsulated ${\rm C}_{60}$ and $^{13}{\rm C}_6{\rm H}_6$ ($^{13}{\rm C}$ enriched benzene). Raman spectroscopy was performed with a Dilor xy triple monochromator spectrometer excited with the lines of an Ar-Kr gas discharge laser. We report herein data obtained with the $514.5\,\rm nm$ line of the Ar ion only, as we focus on the energetic 2D Raman line of the inner 
tubes which is strongest for this excitation energy \cite{Pfeiffer2005PRB}.

We performed first order Raman calculations with the semi-empirical PM3 method as implemented in the Gaussian09 package \cite{g09}. We neglected the outer tube and the inner tube was treated as a molecule: a hydrogen-terminated piece of (5,5) armchair type SWCNT consisting of 600 carbon and 20 hydrogen atoms. The structure was first fully relaxed with {\em opt=tight} option, then we obtained the force constants and the polarizability derivatives. The Raman activities for different distribution of isotope masses were evaluated using the {\em freqchk} utility of Gaussian09 \cite{g09}. We calculated the simulated Raman spectra using the GaussSum tool with a laser energy of $2.4\,\rm eV$ and linewidth of $1\,\rm cm^{-1}$ at room temperature \cite{gausssum}. Finally, averaging of Raman spectra was performed on 2000 random isotope configurations with the exception of the clusters of 30 carbon atoms, where all 756 possible configurations were taken into account. We compared the Raman spectra of small molecules (
methane, 
benzene, ${\rm C}_{60}$ fullerene) calculated with the PM3 method to the first principles based (DFT/B3LYP) results and found that while the frequencies obtained by the semi-empirical method can be very different from the first principles results, the Raman activities of the more sophisticated method are very well reproduced.

\section{Results and discussion}

\begin{figure}
\centering
\includegraphics[width=0.9\hsize]{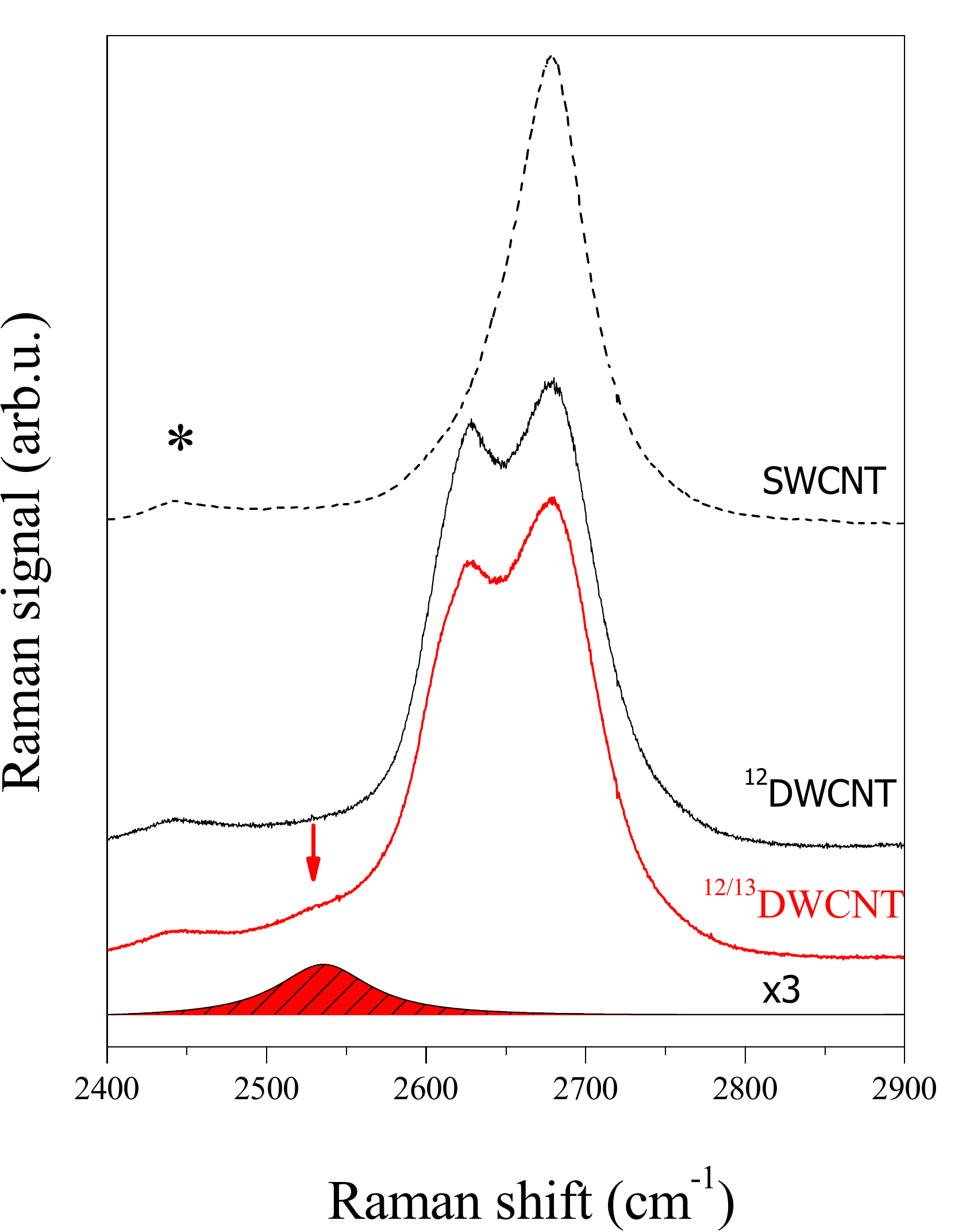}
\caption{The 2D Raman line spectral range for the $^{12}$DWCNT and $^{12/13}{\rm DWCNT}$ samples. Dashed curve is the 2D Raman line for the starting SWCNT sample. Asterisk shows a mode which is present for all samples. Arrow indicates the line which is significantly downshifted in the $^{12/13}{\rm DWCNT}$ sample; a shaded area visualizes the shape of the mode.}
\label{Fig:Fig1_DWCNT_general}
\end{figure}

In \ref{Fig:Fig1_DWCNT_general}, we show the 2D Raman line range for $^{12}{\rm DWCNT}$ and $^{12/13}{\rm DWCNT}$ samples. The spectrum for a single-wall carbon nanotube is shown for comparison. The double peak structure with lines at 2624(1) and 2680(1) (with 25 and 33 $\rm cm^{-1}$ linewidths) was identified previously\cite{Pfeiffer2005PRB} as 2D Raman lines coming from the inner and outer walls, respectively. We note that the inner tube intensity is smaller herein than for purely ${\rm C}_{60}$ based inner tubes. This is caused by the reduced amount of carbon when benzene is co-encapsulated with ${\rm C}_{60}$ inside the host tubes. It was shown \cite{SimonCPL2006} that benzene, which contains one tenth of carbon compared to ${\rm C}_{60}$ per molecule, occupies about the same volume as ${\rm C}_{60}$, thus there is less room for the latter and as a result there is less carbon available for the growth of the inner tubes.

The choice of the 2D Raman line for an experimental study of the isotope shift is supported by the following arguments: i) the inner and outer tube 2D components are split by about $50\,\rm cm^{-1}$ whereas e.g. the inner and outer tube G modes overlap within $5\,\rm cm^{-1}$ \cite{SimonPRL2005}, ii) beside of the inner-outer tube splitting, it has no structure and the 2D Raman lines are nearly featureless Lorentzians for natural carbon SWCNTs (this contrasts to e.g. the LO-TO splitting of the G mode), iii) the high energy of the 2D line enhances the isotope induced shifts, iv) the inner tube 2D Raman line is in fact stronger than that of the outer tube 2D Raman line \emph{per carbon amount} particularly at $514.5\,\rm nm$ excitation \cite{Pfeiffer2005PRB}.

$^{13}{\rm C}$ isotope enrichment of carbon nanotubes induces a downshift of the corresponding Raman modes and a broadening due to the inhomogeneous distribution of the isotopes \cite{SimonPRL2005,ZolyomiPRB2007}. The overall downshift of the Raman modes corresponds to the average amount of carbon isotopes according to:

\begin{equation}
\frac{f}{f_0}=\left(\frac{12.011+c\cdot 13}{12.011}\right)^{1/2}
\label{eq:downshift}
\end{equation}

\noindent where $f$ and $f_0$ are the Raman shifts with and without $^{13}{\rm C}$ doping, $12.011\,\rm g/mol$ is the molar mass of natural carbon and it reflects the $1.1\%$ abundance of $^{13}{\rm C}$ in natural carbon. The broadening of Raman lines has a maximum for $50\%$ enrichment where the statistical randomness of the two kinds of isotopes is largest. 

The same effect, i.e. a simultaneous line shifting and broadening was expected herein for the inner tubes grown from a mixture of $^{13}{\rm C}_6{\rm H}_6$ and ${\rm C}_{60}$. To our surprise, the inner tube 2D Raman line does not shift to lower Raman frequencies in an overall manner for the $^{12/13}{\rm DWCNT}$ sample, but instead a well defined peak develops as a shoulder on the low frequency side of the inner tube 2D Raman line (arrow in \ref{Fig:Fig1_DWCNT_general}).

\begin{figure}
\centering
\includegraphics[width=0.9\hsize]{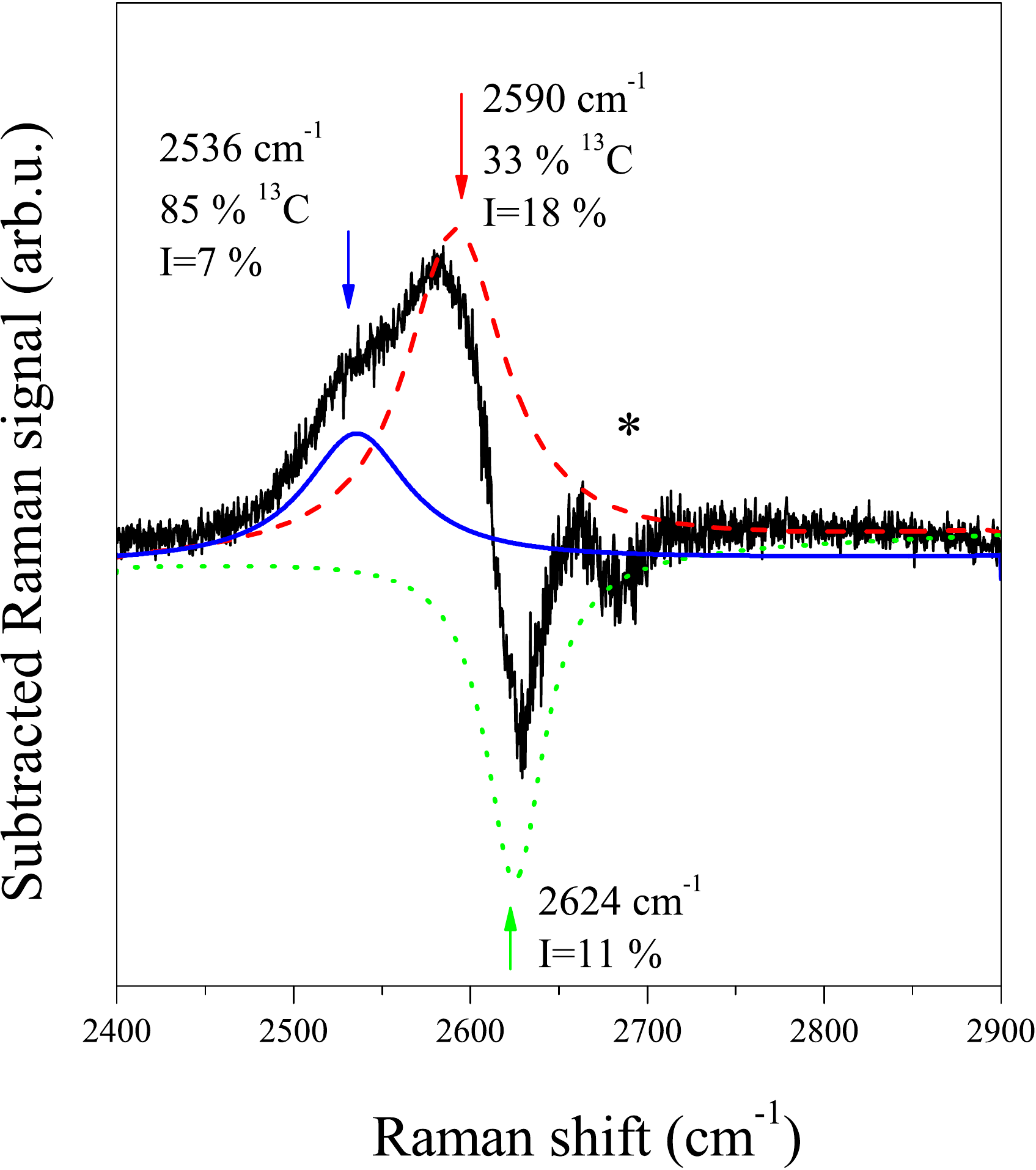}
\caption{Spectrum obtained after subtracting the 2D Raman line of the $^{12}{\rm DWCNT}$ sample from that of the $^{12/13}{\rm DWCNT}$. A deconvolution into several components is also shown. The Raman shifts, the nominal $^{13}{\rm C}$ enrichment level and the intensity of the particular component with respect to the inner tube 2D Raman line are shown.}
\label{Fig:Fig2_13DWCNT_subtracted}
\end{figure}

In order to substantiate the appearance of the unexpected feature in the 2D Raman line of the $^{12/13}{\rm DWCNT}$ sample, in \ref{Fig:Fig2_13DWCNT_subtracted} we show the spectrum obtained after subtracting the 2D Raman line of the $^{12}{\rm DWCNT}$ sample from the data on the $^{12/13}{\rm DWCNT}$ sample. Four distinct features are observed after the subtraction: i) there is some residual feature near the Raman shift of the outer tube 2D Raman line (asterisk in \ref{Fig:Fig2_13DWCNT_subtracted}), ii) there is a negative component at $2624\,\rm cm^{-1}$ (dotted curve in \ref{Fig:Fig2_13DWCNT_subtracted}), this means that some spectral weight is shifted from the $^{12}{\rm C}$ inner tube toward lower Raman shifts, iii) there is a component at $2590\,\rm cm^{-1}$ (dashed curve in \ref{Fig:Fig2_13DWCNT_subtracted}) which corresponds to an effective $^{13}{\rm C}$ enrichment of $33\%$ an it accounts for $18\%$ of the intensity of the inner tube signal in the $^{12}{\rm DWCNT}$ sample, and iv) there is a 
component at $2536\,\rm cm^{-1}$ (solid curve in \ref{Fig:Fig2_13DWCNT_subtracted}) which corresponds to an effective $^{13}{\rm C}$ enrichment of $85\%$ and an intensity of $7\%$ relative to the inner tube signal intensity in the $^{12}{\rm DWCNT}$ sample.

Observation i) is due to an interaction between the inner and outer tubes and is also  observed without isotope enrichment. Point ii) is a natural consequence of the isotope enrichment: with the down shifting of the Raman lines, the spectral weight is missing from the original position. Points ii) and iii) are surprising observations as they indicate a higher than expected (nominally $9\%$) enrichment levels. We know from previous studies \cite{SimonCPL2006} that this effect cannot be explained by e.g. a higher fraction of benzene molecules. A natural suggestion is that $^{13}\rm C$ rich regions are formed which is embedded in a $^{12}\rm C$ rich matrix rather than observing a fully random mixture of the two isotopes. An additional observation is that no strict conservation of the Raman line intensity applies as indicated by the intensities in \ref{Fig:Fig2_13DWCNT_subtracted}.

\begin{figure}
\centering
\includegraphics[width=0.9\hsize]{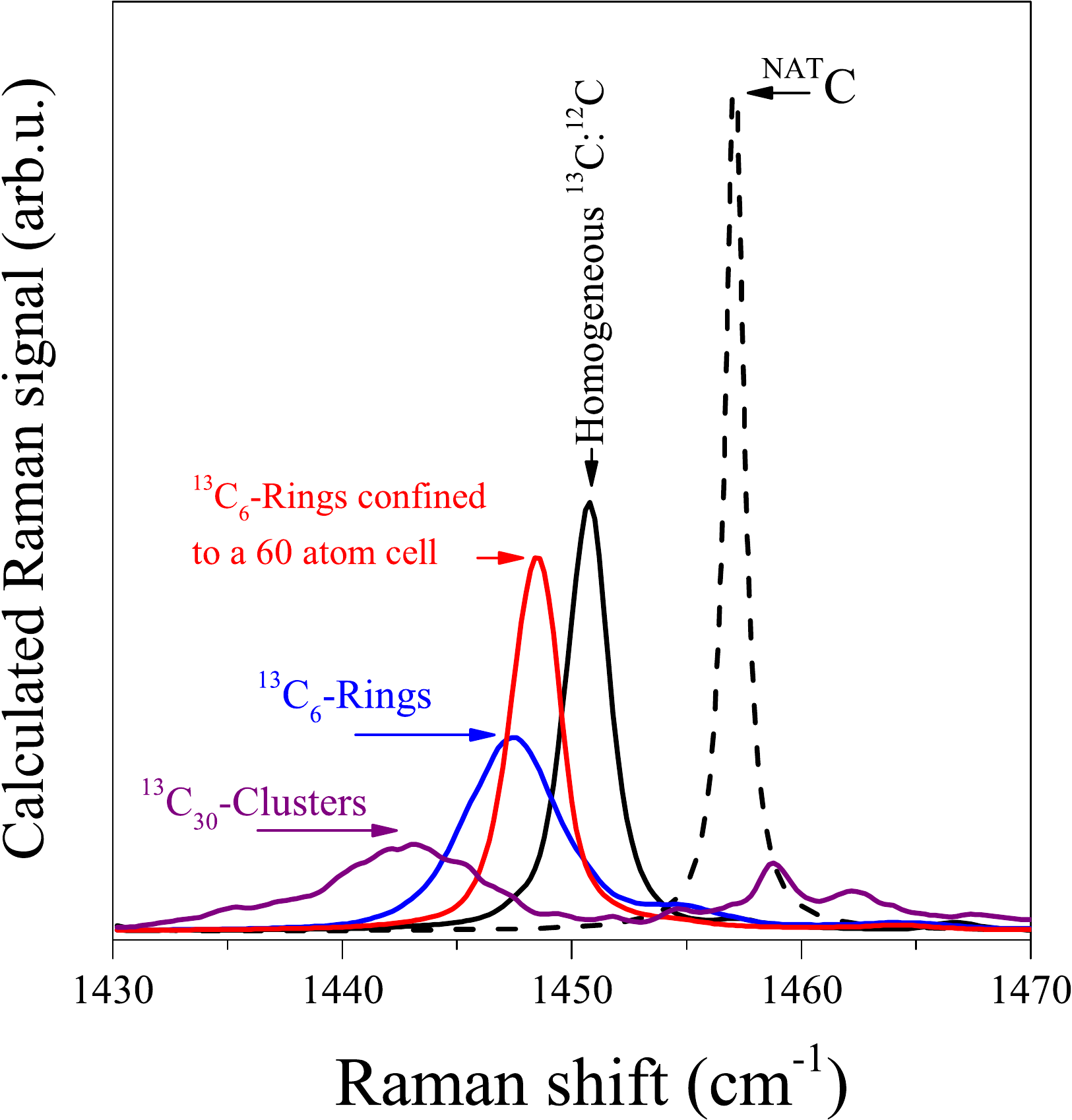}
\caption{Averaged simulated Raman  spectra for natural carbon and for four different types of clusterization in the G band region. The figure confirms that upon a higher level of clusterization the downshift of the Raman peaks can be higher than the real substitution ratio (see text).}
\label{Fig:Fig3_simRaman}
\end{figure}

In \ref{Fig:Fig3_simRaman}, we show the averaged simulated Raman spectra of a SWCNT around the G peak region. We will discuss below the reasons for studying theoretically only the most intense, first order peak. The black dashed curve shows the Raman spectra of the natural case, where all carbon atoms have the same atomic mass (12.011). Note that the position of the G peak ($1457\,\rm cm^{-1}$) is smaller than expected. This is due to the applied semi-empirical method. In principle one could rescale the force constants to fit the experimental value, but we do not bother the absolute position of the calculated Raman peaks, because we always use the same force constant matrix and we only compare the isotope effect based shifts via an effective $^{13}{\rm C}$ enrichment.

The solid black curve shows the homogeneous case, when we replace $10\%$ (60 of the 600) carbon atoms with $^{13}{\rm C}$ atoms in a completely random manner. We note that this enrichment level is not fully what is used in the experiment (60 $^{13}\rm C$ atoms out of 660) but this choice simplifies the calculation, which is performed for a qualitative demonstration. The G peak is downshifted by $10\%$ according to eq. (\ref{eq:downshift}) and broadened in agreement to our previous result \cite{ZolyomiPRB2007}. The red and blue 
curves show the cases where we substitute 10 rings of 6 atoms (in a hexagon, i.e. the carbon atoms of a benzene), the only difference being that for the blue curve we choose the hexagons in a completely random manner, while for the red curve we add an extra restriction: we select one hexagon in every 60 atom segment of the SWCNT. In both cases $10\%$ of the carbon atoms are $^{13}{\rm C}$ substituted, but the effective $^{13}{\rm C}$ enrichment (as obtained from the Raman line downshift) is $16\%$ and $14\%$, respectively. Moreover the restriction leads to a smaller broadening. Finally, the purple curve shows the case where the clusterization is larger: two clusters of 30 atoms on a cylinder (1.5 unit cells of the (5,5) SWCNT) are substituted by $^{13}{\rm C}$. In this case the overall substitution is again $10\%$, but the effective $^{13}{\rm C}$ enrichment according to the downshift of the most prominent peak is $23\%$. The blushifted peak at $\sim 1459\,\rm cm^{-1}$ will be discussed below.

Our theoretical finding supports the experimental observation, that upon a higher level of clusterization the downshift of the Raman peaks can be higher than the real substitution ratio. However, we could not reproduce the most significantly downshifted component. This might be due to the specific character of the 2D Raman line \cite{thomsen_double_2000}. The 2D Raman line originates from phonons at the edges of the Brillouin zone, being more spread in \textbf{k}-space, they are more localized in real space, therefore they are expected to be more sensitive to clusterization. A similar analysis for the 2D Raman line would be computationally extremely demanding, since there is an additional integration over \textbf{k}-space for the 2D Raman line. This explains why we chose a conventional $\Gamma$ point phonon for the calculations, i.e. the G mode, whereas we analyzed the 2D Raman line experimentally due to the reasons described above.

The effect of $^{12}{\rm C} \rightarrow {^{13}{\rm C}}$ substitution on the Raman spectra is twofold. The frequency of the modes are always downshifted, but the value of the downshift varies depending on the positions of the substituted carbon atoms. Furthermore, it affects the Raman intensity of the peaks: the magnitude of the peak can decrease or increase drastically, so there is no strict conservation of Raman intensities in agreement with the experimental observation. In some cases, this can be seen as an ``apparent blueshift'' \cite{kurtipssb2015}, which explains the peak at $\sim 1459\,\rm cm^{-1}$ for the $^{13}{\rm C}_{30}$-clusters in \ref{Fig:Fig3_simRaman} (purple curve).

\section{Conclusions}
In summary, we studied a compelling isotope engineered carbon material: single-wall carbon nanotubes which are grown from $^{13}{\rm C}$ enriched benzene and natural carbon containing ${\rm C}_{60}$. We observe an unexpected downshift of the Raman modes, which is best visible for the 2D Raman line: the downshift is much larger than expected for a homogeneously distributed isotope enrichment. A semi-empirical based modeling of the Raman mode energies for $^{13}{\rm C}$ isotope enrichment suggests that the experimental observation is compatible with a significant clustering of the isotopes. We believe that our observation opens the way for further combined molecular-isotope engineering, i.e. when several different molecules are filled in the host nanotubes as inner tube precursors with different levels of isotopes. This could result in an inner tube whose isotope distribution is controlled to a high degree.

\section*{Acknowledgements}
The Hungarian National Research, Development and Innovation Office (NKFIH) Grant Nr.~K108676, K115608 and K119442 and the European Research Council Starting Grant Nr.~ERC-259374-Sylo are acknowledged for support.

\bibliography{LinArrayC13}

\begin{tocentry}
\includegraphics{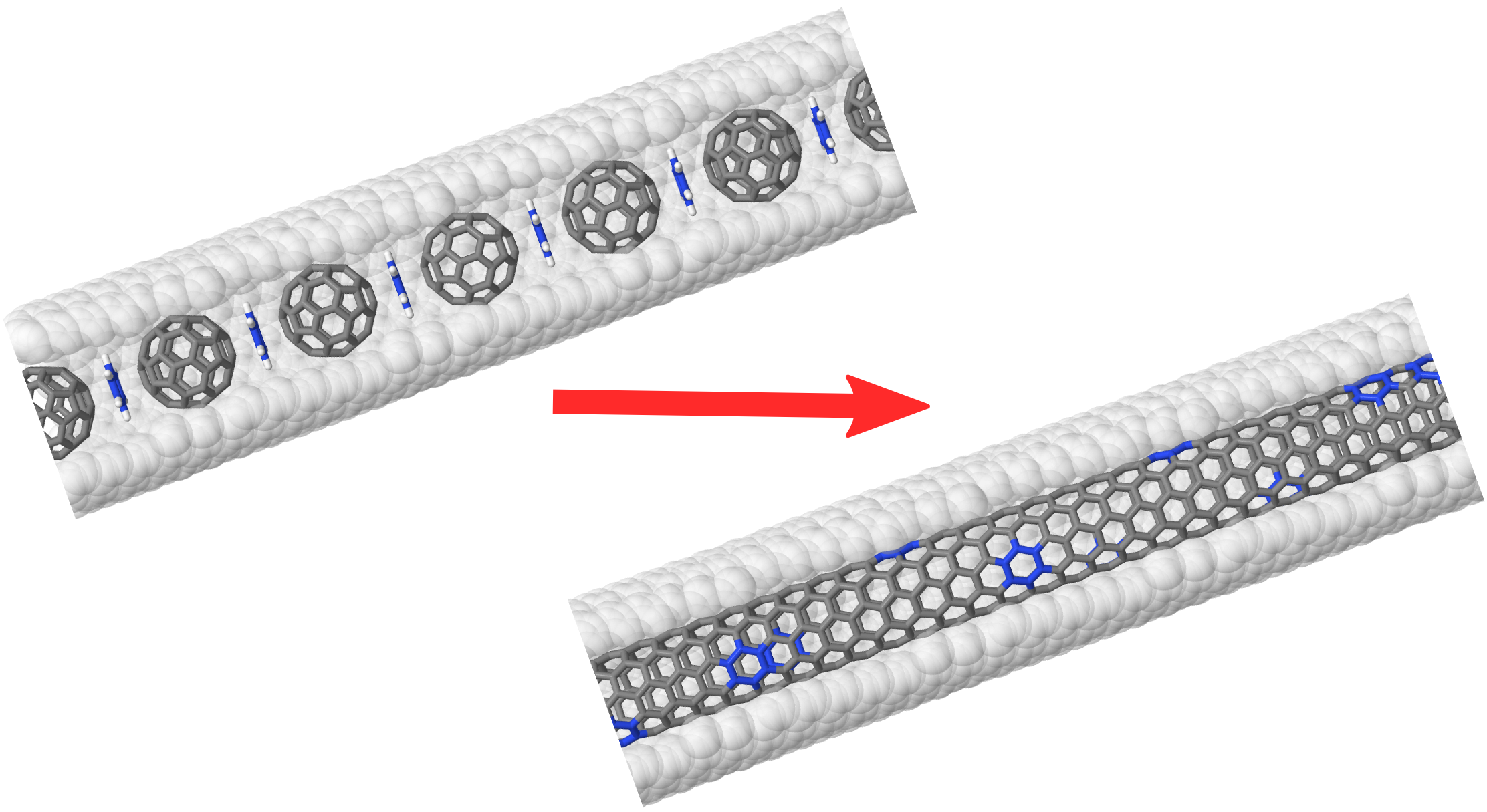}
\end{tocentry}

\end{document}